\documentclass[11pt]{article}

\def\be{\begin{equation}}
\def\ee{\end{equation}}
\def\bea{\begin{eqnarray}}
\def\eea{\end{eqnarray}}
\def\p{\partial}
\def\bp{\bar{\partial}}
\def\nn{\nonumber \\}

\title{On the stability and spectrum of non-supersymmetric $AdS_5$
  solutions of M-theory compactified on K\"ahler-Einstein spaces} 

\author{
Jonathan E. Martin\\School of Physics and Astronomy, University of Nottingham, NG7 2RD, UK\\ppxjm1@nottingham.ac.uk\\
\\
Harvey S. Reall\\Department of Applied Mathematics and Theoretical Physics,\\Centre for Mathematical Sciences, Wilberforce Road, Cambridge, CB3 0WA, UK\\hsr1000@cam.ac.uk}

\begin{document}

\maketitle

\begin{abstract}
Eleven-dimensional supergravity admits non-supersymmetric solutions of
the form $AdS_5 \times M_6$ where $M_6$ is a positive K\"ahler-Einstein space. We show
that the necessary and sufficient condition for such solutions to be
stable against linearized bosonic supergravity perturbations can be
expressed as a condition on the spectrum of the Laplacian acting on
$(1,1)$-forms on $M_6$.  For $M_6=CP^3$, this condition is satisfied,
although there are scalars saturating the Breitenl\"ohner-Freedman
bound. If $M_6$ is a product $S^2 \times M_4$ (where $M_4$ is
K\"ahler-Einstein) then there is an instability if $M_4$ has a
continuous isometry.
We show that a potential non-perturbative instability due to 5-brane
nucleation does not occur. The bosonic Kaluza-Klein spectrum is
determined in terms of eigenvalues of operators on $M_6$. 
\end{abstract}

\section{Introduction}
Eleven-dimensional supergravity admits well-known ``Freund-Rubin''
compactifications of the form $AdS_4 \times M_7$  or
$AdS_7 \times  M_4$, where $M_7$ and $M_4$ are positive Einstein
manifolds \cite{kkreview}. Less well-known is the fact that there are also solutions
of the form $AdS_5 \times M_6$ where $M_6$ is a six dimensional positive
K\"ahler-Einstein space \cite{dolan84a}. The solutions have metric\footnote{
Our conventions are summarized in Appendix A.}
\be
 ds^2 = g_{\mu \nu}(x) dx^\mu dx^\nu + g_{mn}(y)dy^m dy^n,
\ee
where $g_{\mu\nu}$ and $g_{mn}$ are the metrics on $AdS_5$ and $M_6$
respectively,  with Ricci tensors 
\be
 R_{\mu\nu} = -2c^2 g_{\mu\nu}, \qquad R_{mn} = 2c^2 g_{mn},
\ee
so the radius of $AdS_5$ is $\ell = \sqrt{2}/|c|$. The 4-form is
\be
 F = c J \wedge J,
\ee
where $J$ is the K\"ahler form on $M_6$. Examples of suitable $M_6$
are: $CP^3$; the quotient $SU(3)/T$ where $T$ is the maximal torus
of $SU(3)$; the Grassmanian $Gr_2(R^5)$; or a product\footnote{
In the case in which $M_6$ is a product of lower-dimensional
K\"ahler-Einstein spaces, i.e., $M_6 = M_4 \times S^2$, these
solutions can be generalized by taking $F = c_4 J^{(4)} \wedge J^{(4)}
+ c_2 J^{(4)} \wedge J^{(2)}$, where $J^{(4)}$, $J^{(2)}$ are the
K\"ahler forms on $M_4$ and $S^2$ respectively. This gives a
2-parameter family of solutions with independent radii for $M_4$ and $S_2$
\cite{dolan84}. Similarly, if $M_6 = S^2 \times S^2 \times S^2$ then
one can obtain a 3-parameter family.
We shall not consider these generalizations further.}
$M_4 \times S^2$ where the only possible $M_4$ are
$CP^2$, $S^2 \times S^2$, or a del Pezzo surface $dP_k$, $k=3\ldots 8$ \cite{tian,tianyau}. 
This list includes all cases for which $M_6$ is either homogeneous or
a product (inhomogeneous non-product $M_6$ also exist \cite{tian}). These solutions are
not supersymmetric: for $M_6=CP^3$ this was proved in \cite{popevann},
and for general $M_6$ it follows from the analysis of supersymmetric
$AdS_5$ solutions of \cite{gauntlett}. 

By the AdS/CFT correspondence \cite{adscftreview}, these solutions should be dual to conformal field theories in four dimensions. Flux quantization renders $c$ discrete. For $M_6=CP^3$, 
the central charge of the CFT dual to these solutions scales as $N^3$, 
where $N$ is the number of units of flux on $CP^2 \subset CP^3$
\cite{adscftreview}. This suggests that these solutions may have an
interpretation in terms of M5-branes wrapping a 2 cycle. The supergravity approximation is valid for large $N$.

The purpose of this
paper is to investigate the stability of these solutions.
We shall examine three potential instabilites. First,
we check whether there is a non-perturbative instability due to
quantum nucleation of M5-branes (wrapping a 2-cycle in $M_6$) 
\cite{dowker,seiberg}. We find that this does not happen for
any $M_6$: the 5-brane (Euclidean) action is always positive 
and an instanton describing such a process never exists.

Secondly, we consider perturbative stability by considering linearized
supergravity perturbations. We determine the full bosonic Kaluza-Klein
(KK) spectrum for general $M_6$ in terms of eigenvalues of
differential operators on $M_6$. The gauge group is $G \times U(1)^{b_2-1}$, where
$G$ is the isometry group of $M_6$ and $b_2$ the second Betti number
of $M_6$. The squared masses of all fields are
non-negative except possibly for scalars arising from $(1,1)$-forms on
$M_6$. Demanding that such modes respect the Breitenl\"ohner-Freedman (BF)
stability bound \cite{bf} gives a criterion for stability of these
solutions valid for general $M_6$. Analogous results for Freund-Rubin
compactifications of the form $AdS_4 \times M_7$ were obtained in
\cite{duffpopestability}, and for Freund-Rubin compactifications of
other theories in \cite{freedmangubser}.

Our criterion is as follows. Consider transverse,
primitive,\footnote{``Primitive'' means that the contraction with the
K\"ahler form vanishes.} $(1,1)$-form eigenfunctions of the Hodge-de
Rham Laplacian on $M_6$ with eigenvalue $\lambda_{(1,1)}$. A K\"ahler-Einstein
compactification $AdS_5 \times M_6$ suffers a linearized bosonic {\it
  instability} if, and only if, there is a mode with
\be
 2c^2 < \lambda_{(1,1)} < 6c^2.
\ee
We have investigated the spectrum for some of the $M_6$ listed
above. The results are given in table \ref{resultstable}.
\begin{table}
\begin{center}
\begin{tabular}{|l|l|l|l}
\hline
$M_6$ & Isometry group & Classically stable? \\
\hline
$CP^3$ & $SU(4)$ & yes \\
$S^2 \times S^2 \times S^2$ & $SO(3)^3$ & no \\
$S^2 \times CP^2$ & $SO(3) \times SU(3)$ & no \\
$S^2 \times dP_3$ & $SO(3) \times U(1)^2$ & no \\
$S^2 \times dP_{k>3}$ & $SO(3)$ & ? \\
$SU(3)/T$ & $SU(3)$ & ? \\
$Gr_2(R^5)$ & $SO(5)$ & ? \\
\ldots & \ldots & \ldots \\
\hline
\end{tabular}
\end{center}
\caption{Classical linearized stability results for particular $M_6$}
\label{resultstable}
\end{table}
For $M_6=CP^3$, the lowest eigenvalue is $\lambda_{(1,1)} =
6c^2$. Hence $AdS_5 \times CP^3$
is stable at the linearized level in classical supergravity. However,
there are scalar fields that saturate the BF bound. Therefore an
analysis of finite $N$ corrections to the mass would be required to make
a definite statement about perturbative stability.\footnote{
These corrections are of two types. Higher derivative corrections give contributions scaling as powers of $1/N$. Quantum loop corrections give contributions scaling as powers of $1/N^3$.} The scalars saturating the bound transform in the $[0,2,0]$ representation of $SU(4)$.

For $M_6 = S^2 \times M_4$, one might expect an instability
corresponding to the $S^2$ increasing in radius and $M_4$ decreasing (or vice versa)
since this is what happens for product space Freund-Rubin
compactifications \cite{duffpopestability}. However, such a mode
corresponds to $\lambda_{(1,1)}=0$, and is therefore stable: the
flux on the internal space stabilizes the solution against this kind
of deformation. However, we find that there is a mode with
$\lambda_{(1,1)}=4c^2$ whenever $M_4$ possesses a continuous
isometry. This implies that $S^2 \times S^2 \times S^2$, $S^2 \times
CP^2$ and $S^2 \times dP_3$ give unstable solutions. However $dP_k$
has no continuous isometries for $k>3$ \cite{james}, so the classical 
stability of $S^2 \times dP_k$ for $k>3$ requires further investigation.

It would be interesting to determine the spectrum for the other
homogeneous spaces $Gr_2(R^5)$ and $SU(3)/T$. We note that
$SU(3)/T$ possesses a primitive harmonic $(1,1)$-form, so
the lowest eigenvalue is $\lambda_{(1,1)}=0$ in this case, as for the
product spaces.

The third possible instability that we have considered is the possibility that quantum corrections could generate a tadpole for a massless, uncharged, scalar field, resulting in runaway behaviour \cite{berkooz}. 
To examine this possibility, we need to investigate whether
there are massless scalars transforming as singlets under $G$ (as no
fields are charged under $U(1)^{b_2-1}$).

A massless scalar will be present if $M_6$ admits complex
structure moduli. Now, $dP_k$ has such moduli for $k>4$
\cite{james}. Hence $M_6 = S^2 \times dP_k$ has such moduli.
These are trivially singlets under $G$ (since $dP_k$ has
no continuous symmetries for $k \ge 4$). Therefore we conclude that no symmetry
prevents quantum
corrections from destabilizing compactifications with $M_6 =S^2 \times
dP_k$ for $k>4$, at least at a generic point in moduli space (at
special points there may be discrete symmetries preventing this from
happening). Clearly this can happen whenever $M_6$ has complex
structure moduli invariant under $G$, in particular if $M_6$ has
complex structure moduli and no isometries.

This paper is organized as follows. In section 2, we give a detailed
summary of our results. We first investigate quantum nucleation of
M5-branes. We then summarize our analysis of the Kaluza-Klein
spectrum, explain the origin of our stability criterion, and
investigate this criterion for several possible $M_6$. Section 3
contains the full calculation of the Kaluza-Klein spectrum.

\section{Results}

\subsection{5-brane nucleation}

A potential non-perturbative instability involves quantum nucleation
of branes \cite{dowker,seiberg}. Since the solutions are purely magnetic, we
need only consider nucleation of 5-branes. The (Euclidean) 5-brane action is
\be
 S = T  \int d^6 \xi \sqrt{h} - T \int C_{(6)},
\ee
where $T$ is the 5-brane tension, $\xi$ are worldvolume coordinates,
$h$ the determinant of the induced metric on the worldvolume and
$C_{(6)}$ the 6-form potential for $\star F$.\footnote{Note that the
  M-theory Chern-Simons term vanishes for these solutions so there are
  no subtleties in defining $C_{(6)}$.} For the solutions of interest,
$\star F = 2c \eta_5 \wedge J$, where $\eta_5$ is the volume form of
$AdS_5$. We are looking for instanton solutions so we work in
Euclidean signature, writing the metric on Euclidean $AdS_5$ as 
\be
 ds^2 = d\rho^2 + \ell^2 \sinh^2(\rho/\ell) d\Omega_4^2.
\ee
We can choose the gauge ($\ell = \sqrt{2}/|c|$)
\be
 C_{(6)} = \frac{8}{c^3} \left[ \int_0^\rho \sinh^4 (c x /\sqrt{2}) dx
   \right] d\Omega_4 \wedge J. 
\ee 
To get a non-trivial contribution from the flux term in the action, we
take the 5-brane  worldvolume to be $S^4 \times \Sigma$ where $S^4$ is
a sphere of constant $\rho$ in $AdS_5$ and $\Sigma$ a 2-cycle in
$M_6$. Upon continuing to Lorentzian signature this would give an
exponentially expanding 5-brane with worldvolume $dS_4 \times
\Sigma$. Evaluating the action on this Ansatz gives 
\be
 S = \frac{4T}{c^4} \Omega_4 \left[\sinh^4 (c \rho/\sqrt{2}) V - 2c
   \int_0^\rho \sinh^4 (c x /\sqrt{2}) dx \, \int_\Sigma J \right], 
\ee
where $V$ is the volume of $\Sigma$. Varying with respect to $\rho$
gives the condition for a turning point (for $c>0$) 
\be
 \tanh (c \rho/\sqrt{2}) = \frac{\sqrt{2} V}{\int_\Sigma J} \ge \sqrt{2},
\ee
where the inequality follows from the fact that $J$ is a calibration
in $M_6$. Hence there is no solution for $\rho$ (the action is
positive and monotonically increasing with $\rho$) so we conclude that
there is no 5-brane nucleation instability. 

It would be interesting to investigate more complicated
non-perturbative instabilities, such as the one of \cite{horowitz},
which involves simultaneous nucleation of branes and a Kaluza-Klein
bubble. However, since $M_6$ must be simply connected \cite{besse}, our
spacetimes do not contain a circle that can collapse to zero size at a
bubble. Perhaps there could be an instability involving a bubble
describing the collapse of a higher-dimensional submanifold of
spacetime, e.g.  an $S^2$ inside $M_6$.

\subsection{The Kaluza-Klein spectrum}

\subsubsection{Harmonics on $M_6$} 

To determine the KK spectrum, we expand each field in terms of
harmonics on $M_6$. These harmonics satisfy various conditions. In
particular, we will be concerned with $(p,q)$-form eigenfunctions of
the Hodge-de Rham Laplacian
\be
 \Delta_6 \hat{Y}_{(p,q)} = \lambda_{(p,q)} \hat{Y}_{(p,q)},
\ee
which are {\it primitive}:
\be
 J^{mn} \hat{Y}_{(p,q)mn\ldots} = 0,
\ee
and {\it transverse}:
\be
 d_6^\dagger \hat{Y}_{(p,q)} = 0.
\ee
A hat on a $(p,q)$-form will be used to denote that it is primitive
and transverse. As we explain below, a general $(p,q)$-form can be
decomposed into a primitive, transverse piece and pieces built from
forms of lower rank.

For $CP^N$, the spectrum of the Laplacian acting on $(p,q)$
forms was determined in \cite{cp3eigenvalues}. Using these results, one can determine the eigenvalues of the Laplacian acting on transverse primitive forms on $CP^3$. These are summarized in table \ref{cp3eigenvalues}.
\begin{table}
\begin{center}
\begin{tabular}{|c|c|c|c|c|c|}
\hline
 $\lambda$ & $\lambda_{(1)}$ & $\lambda_{(1,1)}$ &$\lambda_{(2,0)}$ & $\lambda_{(2,1)}$  \\
\hline
 $c^2 k(k+3)$ & $c^2(k+2)(k+4)$ & $c^2 (k+2)(k+3)$ & $c^2 (k+3)(k+4)$ & $c^2 (k+2)(k+4)$ \\
\hline
$[k,0,k]$ & $[k,1,k+2]$ & $[k,2,k]$ & $[k,0,k+4]$ & $[k,1,k+2]$  \\
\hline
\end{tabular}
\end{center}
\caption{Eigenvalues of the Laplacian on $CP^3$ acting on transverse primitive forms, determined from \cite{cp3eigenvalues}. $k$ is a non-negative integer. $\lambda \equiv \lambda_{(0,0)}$, $\lambda_{(1)} \equiv \lambda_{(1,0)}$. There are no transverse $(3,0)$-forms. The bottom row gives the corresponding representation of $SU(4)$. If a $(p,q)$-form eigenfunction belongs to representation $[r,s,t]$ then the $(q,p)$-form eigenfunction belongs to the complex conjugate representation $[t,s,r]$.}
\label{cp3eigenvalues}
\end{table}

We recall a few facts about eigenfunctions of the Hodge-de Rham
Laplacian on general $M_6$ \cite{besse,pope}. There are no harmonic
$(p,0)$-forms so $\lambda_{(p,0)} = \lambda_{(0,p)}>0$. In particular,
this implies there are no harmonic 1-forms. It also implies that there are no transverse $(3,q)$-forms since such forms would be annihilated by both $\partial$ and $\partial^\dagger$, and hence be harmonic.
For scalars, which we shall take to be real, 
non-constant eigenfunctions have $\lambda \ge 4c^2$. 
Eigenfunctions saturating the bound are in one-to-one
correspondence with Killing vector fields. This is because a vector
field $V$ on $M_6$ is Killing if, and only if, it can be written as
$d_6^c Y$ where $Y$ is a scalar eigenfunction with $\lambda=4c^2$. 

\subsubsection{$(1,1)$-form perturbations}\label{sec:s-1,1f}

We perform a full linearized analysis of the bosonic
Kaluza-Klein spectrum in section \ref{sec:spectrum}. The result of
this analysis is that the only modes that could violate the
Breitenl\"ohner-Freedman stability bound, indeed the only modes with
negative squared mass, arise from $(1,1)$-forms on $M_6$. These
are associated with hermitian metric perturbations on $M_6$ (i.e. perturbations
for which, in complex coordinates, the $zz$ and $\bar{z} \bar{z}$
components of the metric perturbation vanish). 
Explicitly, the metric perturbation is
\be
 \delta g_{mn}(x,y) = -\sum_I h^I(x) \hat{Y}^I_{(1,1)mp}(y) J^p{}_n.
\ee
Here we have performed the usual Kaluza-Klein decomposition of modes
into a product of fields in $AdS_5$ and $M_6$. The former are the
scalars $h^I(x)$. On $M_6$, $\hat{Y}^I_{(1,1)}$ denotes a
primitive, transverse, $(1,1)$-form eigenfunction of the Hodge-de Rham
Laplacian, with eigenvalue $\lambda^I_{(1,1)}$:
\be
 \Delta_6 \hat{Y}^I_{(1,1)} = \lambda^I_{(1,1)} \hat{Y}^I_{(1,1)}.
\ee
Modes with different $I$ will decouple from each other. 
We shall suppress the $I$ index in what follows. 

This metric perturbation will couple to terms in the 4-form
perturbation that also arise from $(1,1)$-forms on $M_6$. These are of
the form
\be
\delta F = d \left(k^-(x) d_6^c \hat{Y}_{(1,1)}(y) \right).
\ee
We can take $\hat{Y}_{(1,1)}$ to be real hence $h$ and $k^-$ are real.

For these modes, the perturbed Maxwell equation reduces to
\be
 (\Delta + \lambda_{(1,1)}) k^- - 4ch = 0 \qquad \lambda_{(1,1)} \ne 0.
\ee
The restriction $\lambda_{(1,1)} \ne 0$ arises from the fact that if
$Y_{(1,1)}$ is harmonic then $d_6^c Y_{(1,1)}$ vanishes hence $k^-$ is
unphysical. The perturbed Einstein equation reduces to
\be
 (\Delta + \lambda_{(1,1)} + 4c^2 ) h - 4 c \lambda_{(1,1)} k^- = 0.
\ee
Hence if $\lambda_{(1,1)}=0$ then we have a single physical real
scalar field $h(x)$ with $m^2 = 4c^2$. 

However, if $\lambda_{(1,1)}>0$ then we have two fields and we need
to diagonalize the above equations to determine the mass spectrum. 
Doing so, we find the masses are given by
\be
 m_\pm^2 = \lambda_{(1,1)} + 2c^2 \pm \sqrt{16 c^2 \lambda_{(1,1)} + 4
   c^4}.
\ee
$m_+^2$ is positive but $m_-^2$ may be negative. An instability occurs if the
Breitenl\"ohner-Freedman bound is violated, i.e.,  $m_-^2 <
-2c^2$. This is equivalent to
\be
\label{eqn:instability}
 2c^2 < \lambda_{(1,1)} < 6c^2 \qquad {\rm for \; \; instability.}
\ee
If there exists a (primitive, transverse) $(1,1)$-form eigenfunction of
the Laplacian on $M_6$ with eigenvalue in this range then the solution
is unstable.

\subsubsection{Stability of $CP^3$}

The results of table \ref{cp3eigenvalues} give
\be
 m_+^2 = c^2 (k+3)(k+6), \qquad m_-^2 = c^2 (k-1)(k+2).
\ee
Hence $m_-^2 \ge -2c^2$ so the Breitenl\"ohner-Freedman bound is
respected. However, modes with $k=0$ give scalar fields that can saturate the bound. These fields transform in the $[0,2,0]$ representation of $SU(4)$. Since there is no supersymmetry to protect the masses, it is necessary to examine whether higher derivative corrections (corresponding to finite $N$ corrections in the dual CFT) raise or lower the masses of these fields in order to make a conclusive statement about perturbative stability.

Note that there are also massless fields arising from modes with $k=1$, in the $[1,2,1]$ of $SU(4)$. Since these are charged under the $SU(4)$ isometry group, a runaway associated with these fields is not expected \cite{berkooz}.

The dimensions of CFT operators dual to the fields arising from $(1,1)$-forms on $CP^3$ are generically irrational (the special $k=0,1$ fields just mentioned excepted).

\subsubsection{Instability of $S^2 \times M_4$}

In Freund-Rubin compactifications, there is generically an instability
if the internal space is a product \cite{duffpopestability,kkreview}. The instability arises from a
metric deformation of the internal space in which one factor in the
product expands and the other contracts. For product space K\"ahler-Einstein
compactifications, we shall see that this simple instability is absent
but there is a more complicated instability, at least if $M_4$ has a
continuous isometry.

Assume that $M_6 = S^2 \times M_4$ where $M_4$ is
K\"ahler-Einstein. The Freund-Rubin product instability arises from
(transverse, traceless) metric perturbations of the form
\be
 \delta g_{mn} \propto h(x) (2 g_{mn}^{(2)} -  g_{mn}^{(4)}),
\ee
where $g_{mn}^{(2,4)}$ are the metrics of $S^2$ and $M_4$
respectively. This is equivalent to a $(1,1)$-form perturbation for
which 
\be
 \hat{Y} \propto 2 J^{(2)} - J^{(4)},
\ee
where $J^{(2,4)}$ are the K\"ahler forms of $S^2$ and $M_4$
respectively (so $J= J^{(2)}+J^{(4)}$). The relative factor in the
above equation is fixed by the primitivity
condition. However, these are covariantly constant hence $\hat{Y}$ is
harmonic, i.e., $\lambda_{(1,1)} = 0$, so these modes do {\it not} lie
within the ``window of instability'' of equation
(\ref{eqn:instability}): they are stable. The presence of flux on the
internal space stabilizes it against this kind of deformation.

To obtain an instability we need to look at more complicated
modes. Consider $M_6 = S^2 \times S^2 \times S^2$ (i.e. $M_4 = S^2
\times S^2$). Let $y_i$ be coordinates, and $J^{(i)}$ the K\"ahler form, of the $i$th
$S^2$. Let $Y$ be a $\lambda=4c^2$ scalar eigenfunction on $S^2$,
which must exist because $S^2$ admits Killing vector fields. Now
consider the following primitive, transverse, $(1,1)$-form on $M_6$:
\be
 \hat{Y} = \left( Y(y_2) - Y(y_3) \right) J^{(1)}(y_1) + \left( Y(y_3)
 - Y(y_1) \right) J^{(2)}(y_2) + \left( Y(y_1) - Y(y_2) \right)
 J^{(3)}(y_3).
\ee
A calculation reveals that this is an eigenfuction of $\Delta_6$ with
eigenvalue $\lambda_{(1,1)} = 4c^2$, i.e., a mode within the range
(\ref{eqn:instability}). Hence $M_6 = S^2 \times S^2 \times S^2$ is an
unstable compactification.

A similar construction works whenever $M_4$ admits a Killing vector
field. Let $Y$ be a scalar harmonic on $M_4$ with eigenvalue
$\lambda$. From this we can build a suitable $(1,1)$-form by 
considering an arbitrary linear combination of $d_4 d_4^c Y$,
$Y J^{(4)}$ and $Y J^{(2)}$ (where $d_4$ is the exterior derivative on
$M_4$), 
and fixing the coefficients by demanding primitivity and
transversality. This gives
\be
 \hat{Y} = d_4 d_4^c Y - \lambda Y J^{(4)} - 2 \lambda Y J^{(2)}.
\ee
This is a $(1,1)$-form eigenfunction of $\Delta_6$ with eigenvalue $\lambda$.
If $M_4$ admits a Killing vector field then there exists a mode with
$\lambda=4c^2$  and hence, from (\ref{eqn:instability}), an
instability. It follows that the $S^2 \times CP^2$ and $S^2 \times dP_3$
are unstable compactifications. However, the K\"ahler-Einstein metric
on $dP_k$ does not admit continuous symmetries for $k>3$ \cite{james}
so we cannot conclude that $S^2 \times dP_k$ is unstable for $k>3$
using this method (unless it could be shown that the lowest
non-trivial eigenfunction of the scalar Laplacian on $dP_k$  has
$\lambda<6c^2$). 

\subsubsection{The full bosonic KK spectrum}

\begin{table}
\begin{center}
\begin{tabular}{| l | l | c | l | l|}
\hline
Field & Type & $m^2$ & Restriction & Section \\
\hline
Spin-2 & real & $\lambda$ &  & \ref{sec:spin-2}\\
\hline
2-form & complex & $\left(\sqrt{\lambda_{(1)} + c^2} \pm c \right)^2$
& $\lambda_{(1)}>0$ & \ref{sec:2f-1f}\\ 
{}&real &  $\lambda + 4c^2$ & $\lambda >0$ & \ref{sec:2f-s}\\
\hline
1-form & complex & $\lambda_{(2,0)}$ & $\lambda_{(2,0)}>0$ & \ref{sec:1f-2f} \\
{}& real & $\lambda_{(1,1)}$ & & \ref{sec:1f-2f}\\
{} & complex & $\left(\sqrt{\lambda_{(1)} + c^2} \pm c \right)^2$ & $\lambda_{(1)}>0$& \ref{sec:1f-1f}\\
 & real & $\lambda+4c^2$ & $\lambda>0$ & \ref{sec:1f-s}\\
 & real & $\lambda+6c^2 \pm \sqrt{(\lambda+6c^2)^2-\lambda(\lambda-4c^2)}$ & only $+$ if $\lambda=0$ & \ref{sec:1f-s}\\
\hline
Scalar & complex & $\lambda_{(2,1)}$ & & \ref{sec:s-3f}\\
& complex & $\lambda_{(2,0)}$ & $\lambda_{(2,0)}>0$ & \ref{sec:s-2f}\\
& real & $\lambda_{(1,1)} +2c^2 \pm \sqrt{16c^2 \lambda_{(1,1)} +
  4c^4}$ & only $+$ if $\lambda_{(1,1)}=0$ & \ref{sec:s-1,1f}\\ 
& complex & $\lambda_{(1,0)}^{(0,1)}$ & & \ref{sec:s-ah}\\
& complex & $\left(\sqrt{\lambda_{(1)} + c^2} \pm c \right)^2$ &
$\lambda_{(1)}>0$ & \ref{sec:s-1f}\\
& real & $\lambda+4c^2$ & $\lambda>0$ & \ref{sec:s-s}\\
& real & $\lambda+6c^2 \pm
\sqrt{(\lambda+6c^2)^2-\lambda(\lambda-4c^2)}$ & only $+$ if
$\lambda=0$ & \ref{sec:s-s}\\
 & real & $0$ (axion) & & \ref{sec:s-s}\\
\hline
\end{tabular}
\end{center}
\caption{The bosonic Kaluza-Klein spectrum. $M_6$ does not admit
  harmonic $(p,0)$-forms, so $\lambda_{(p,0)}>0$. The other restrictions in
  this table arise because the associated modes are unphysical, i.e.,
  give vanishing metric and 4-form perturbations. $\lambda_{(1,0)}^{(0,1)}$ is the eigenvalue of the Laplacian acting on $(1,0)$-forms taking values in the anti-holomorphic tangent space of $M_6$ (which vanishes for infinitesimal complex structure deformations).} 
\label{spectrumtable}
\end{table}

In section \ref{sec:spectrum} we determine the full spectrum of
bosonic KK excitations. The results are summarized in table \ref{spectrumtable}. Note that there
are some curious degeneracies between 2-form, 1-form and scalar
fields. 

For $CP^3$, plugging in the known eigenvalues of the Laplacian acting on $(p,q)$-forms (table \ref{cp3eigenvalues}) gives the mass spectrum of table \ref{cp3spectrum}. The eigenvalue $\lambda_{(1,0)}^{(0,1)}$ can be determined from the eigenvalue of the Lichnerowicz operator acting on anti-hermition tensor modes (see section \ref{sec:s-ah}). The general form of these eigenvalues in terms of a non-negative integer $k$ is known \cite{klr} but the precise lower bound on $k$ is not (i.e. the smallest allowed value of $k$ may be positive).
\begin{table}
\begin{center}
\begin{tabular}{| l | l | c |}
\hline
Field & Type & $m^2/c^2$ ($k=0,1,2,\ldots$) \\
\hline
Spin-2 & real & $k(k+3)$ \\
\hline
2-form & complex & $ (k+2)^2$, $ (k+4)^2$\\
{}&real &  $ (k^2 +5k+8)$\\
\hline
1-form & complex & $ (k+3)(k+4)$\\
{}& real & $ (k+2)(k+3)$\\
{} & complex & $ (k+2)^2$, $ (k+4)^2$\\
 & real & $ (k^2 +5k+8)$\\
 & real & $ k(k+1)$, $ (k+3)(k+4)$\\
\hline
Scalar & complex & $ (k+2)(k+4)$\\
 & complex & $ (k+3)(k+4)$\\
& real & $ (k-1)(k+2)$, $ (k+3)(k+6)$\\
& complex & $(k+1)(k+4)$ \qquad $k \ge \rm{?}$\\
& complex & $ (k+2)^2$, $ (k+4)^2$\\
& real & $ (k^2 +5k+8)$\\
& real & $ k(k+1)$, $ (k+3)(k+4)$\\
 & real & $0$ (axion)\\
\hline
\end{tabular}
\end{center}
\caption{The bosonic Kaluza-Klein spectrum for $CP^3$. The values of $k$ have been shifted to take account of the restrictions in table \ref{spectrumtable}: $k$ is everywhere a non-negative integer except in the row corresponding to $\lambda_{(1,0)}^{(0,1)}$ (see main text).} 
\label{cp3spectrum}
\end{table}

\subsubsection{The massless spectrum}

In addition to the $AdS_5$ graviton, there are massless vector and scalar fields. There is a massless vector for each Killing vector field on $M_6$ (associated with $\lambda=4c^2$ scalar harmonics). These are the usual KK gauge bosons. Massless vectors also arise from primitive harmonic $(1,1)$-forms on $M_6$. These are familiar from Freund-Rubin compactifications \cite{kkreview} except that here we have the primitivity condition. There are $b_2-1$ primitive harmonic $(1,1)$-forms hence the gauge group of the effective 5d theory is $G \times U(1)^{b_2-1}$.

Massless scalar fields need special consideration because, as discussed in the introduction, the presence of uncharged massless scalars may lead to a runaway instability arising from a tadpole generated by quantum corrections \cite{berkooz}.
Massless scalars arise in several ways. First, dualizing the KK zero mode of the M-theory 3-form in $AdS_5$ gives a scalar axion. Classically, this field has a continuous shift symmetry. However, quantum mechanically, the axion may develop a potential generated by M5-brane instantons wrapped on $M_6$. This would break the shift symmetry to a discrete shift symmetry. In either case, the symmetry protects the axion from runaway behaviour.

Second, each Killing field on $M_6$ gives rise to a real massless scalar, which together transform in the adjoint on $G$. If $G$ has rank 3 or greater (i.e. if $M_6$ has at least $U(1)^3$ isometry group - in other words, $M_6$ is toric) then the presence of these scalars can be understood from the fact that solution generating transformations can be used to generate continuous deformations of our background \cite{lunin}. The moduli associated with these deformations correspond to massless scalar fields with exactly flat potentials and these must be at least a subset of the massless scalars arising from Killing fields on $M_6$. If $G$ is simple then it acts transitively on the latter (since they transform in the adjoint of $G$), and hence they must all be moduli. This is the case for $CP^3$. 

If $G$ has an abelian factor then the massless scalar associated with the abelian generator is uncharged hence a runaway is possible. For the spaces listed in table \ref{resultstable}, this happens only for $S^2 \times dP_3$ but we have already seen that this is unstable even at the classical level. 
 
Third, if $M_6$ admits infinitesimal complex structure deformations then these will give
complex massless scalars. These are present e.g. for $S^2 \times dP_{k>4}$ \cite{james}. Since these are uncharged (because $dP_{k>4}$ has no continuous isometries), this suggests that these spaces will indeed suffer from a runaway instability.

Fourth, massless (complex) scalars arise if $M_6$ admits primitive harmonic $(2,1)$-forms. These will be gauge singlets since harmonic forms are invariant under continuous isometries. Hence such scalars could lead to a runaway instability. However, primitive harmonic $(2,1)$-forms do not arise for the spaces listed in table \ref{resultstable}.

Finally, massless scalars arise if there are (transverse, primitive) $(1,1)$-form harmonics with eigenvalue $\lambda_{(1,1)}=12c^2$. One would expect these to be charged in general so they will not generate a runaway. Such scalars are present for $CP^3$ and transform in the $[1,2,1]$ representation of $SU(4)$. 

\section{The Kaluza-Klein spectrum}

\label{sec:spectrum}

\subsection{Decomposition of fields on $M_6$}

On $M_6$, we can decompose a $(p,q)$-form into its primitive part and
a non-primitive part: 
\be
 X_{(p,q)} = X_{0(p,q)} + J \wedge X_{(p-1,q-1)},
\ee
where a subscript $0$ denotes a primitive form. We can further
decompose a primitive form into a transverse part and exact
pieces. Let $\Lambda_0^{(p,q)}$ denote the space of primitive
$(p,q)$-forms. Define a map 
${\cal F}: \Lambda_0^{(p-1,q)} + \Lambda_0^{(p,q-1)} \rightarrow
\Lambda_0^{(p,q)}$ by 
\be
 {\cal F}(Y_{0(p-1,q)}+Z_{0(p,q-1)}) = \left[ \p Y_{0(p-1,q)} + \bp
   Z_{0(p,q-1)} \right]_0,  
\ee
where $[\ldots]_0$ denotes the primitive part.
For given $X_{0(p,q)}$, choose $Y_{0(p-1,q)}$ and $Z_{0(p,q-1)}$ to
minimize the inner product of ${\cal F}(Y_{0(p-1,q)}+Z_{0(p,q-1)})$
with $X_{0(p,q)}$. This results in the orthogonal decomposition 
\be
 X_{0(p,q)} = \hat{X}_{(p,q)} + \left[ \p Y_{0(p-1,q)} + \bp
   Z_{0(p,q-1)} \right]_0, 
\ee
where the hat denotes a form that is both primitive and transverse:
\be
 d^\dagger\hat{X}_{(p,q)}=0 \Leftrightarrow \p^\dagger \hat{X}_{(p,q)}
 = \bp^\dagger \hat{X}_{(p,q)} = 0. 
\ee
For example, we can decompose a general 1-form as
\be
 X_1 = \hat{X}_{(1,0)} + \hat{X}_{(0,1)} + \p X + \bp Y,
\ee
where $X$ and $Y$ are scalars. Using the above decomposition in two
steps shows that 
a general 2-form can be decomposed into terms involving only primitive
transverse forms as 
\be
 X_2 = \hat{X}_{(2,0)} + \hat{X}_{(1,1)} + \hat{X}_{(0,2)} + \p
 \hat{X}_{(1,0)} + \bp \hat{X}_{(0,1)} + \p \hat{Y}_{(0,1)} + \bp
 \hat{Y}_{(1,0)} + \left[ \p \bp Y \right]_0 + J X. 
\ee
To avoid a proliferation of terms, we shall find it more convenient to
work with $n$-forms, rather than $(p,q)$-forms, for most of our
calculations. Note that, in the decomposition of a $n$-form $X_n$ into
$(p,q)$-forms of definite type, the individual $(p,q)$-forms will be
transverse if, and only if, $X_n$ is ``doubly transverse'', i.e., 
\be
 d^\dagger X = d^{c\dagger} X = 0,
\ee
where
\be
 d^c = -i (\p-\bp).
\ee
Hence we can rewrite the 1-form decomposition as (redefining $X$ and $Y$)
\be
 X_1 = \hat{X}_1 + d X + d^c Y,
\ee
and the 2-form decomposition can be rewritten as
\be
 X_2 = \hat{X}_2 + d\hat{X}_1 + d^c \hat{Y}_1 +  d d^c Y  + J X,
\ee
where a hat on a $n$-form denotes that the form is primitive and
doubly transverse. In the penultimate term of the 2-form
decomposition, we have removed the square brackets from $dd^c Y$,
which amounts to shifting the scalar $X$ in the final term. Without
the square brackets, the final two terms are no longer orthogonal but
they are still linearly independent. 

A 3-form $X_3$ can be decomposed as
\be
 X_3 = \hat{X}_3 + d\hat{X}_2 + d^c \hat{Y}_2 + d d^c \hat{X}_1 + J
 \wedge \left( \hat{Y}_1 + d X + d^c Y \right). 
\ee
Now consider a symmetric tensor $h_{mn}$. This can be decomposed into
its hermitian and anti-hermitian parts: 
\be
 h_{mn} = H_{mn} + A_{mn}, \qquad J_m{}^p J_n{}^q H_{pq} = H_{mn},
 \qquad J_m{}^p J_n{}^q A_{pq} = -A_{mn}. 
\ee
The hermitian part is equivalent to a $(1,1)$-form $X$:
\be
 H_{mn} = -X_{(1,1)mp} J^p{}_n.
\ee
$X$ can be decomposed as described above. The anti-hermitian part
$A_{mn}$ can be split into its $(2,0)$ and $(0,2)$ parts. Consider the
map ${\cal F}$ from $(1,0)$-forms to symmetric $(2,0)$ tensors defined
by 
\be
 {\cal F}(X_{(1,0)})_{mn} = \nabla^+_{(m} X_{n)},
\ee
where $\nabla^\pm_m$ denote the projection of of $\nabla_m$ onto its
$(1,0)$ and $(0,1)$ parts respectively. The space of $(2,0)$ symmetric
tensors has the orthogonal decomposition ${\rm Im}(\cal F) + {\rm
  Ker}({\cal F}^\dagger)$ and there is a similar decomposition for
$(0,2)$ symmetric tensors so we can write 
\be
\label{eqn:antihermitian}
 A_{mn} = \hat{A}_{mn} + \nabla^+_{(m} Y_{n)(1,0)} + \nabla^-_{(m} Y_{n)(0,1)},
\ee
where $\hat{A}_{mn}$ is a transverse anti-hermitian tensor:
\be
 \nabla^m \hat{A}_{mn} = 0.
\ee

\subsection{Decomposition of perturbation}

Consider a small perturbation of the solution:
\be
 \delta g_{MN} = h_{MN}, \qquad \delta F_{MNPQ} = f_{MNPQ}.
\ee
The Bianchi identity implies $df = 0$ hence $f=da$ for some 3-form
$a$.\footnote{$M_6$ admits at least one harmonic 4-form (i.e. $J
  \wedge J$) but we assume that $f$ vanishes at infinity in $AdS_5$ so
  we don't need to include a contribution to $f$ proportional to such
  a form.}  

The $AdS_5$ components of the metric perturbation transform as a scalar on $M_6$ and
can be expanded in eigenfunctions of the Hodge-de Rham Laplacian on
$M_6$: 
\be 
 h_{\mu\nu}(x,y) = \sum_I h^I_{\mu\nu}(x) Y^I(y),
\ee
where $\Delta_6 Y^I = \lambda^I Y^I$. Decomposing $h^I_{\mu\nu}$ into
transverse parts gives 
\be
 h_{\mu\nu}(x,y) = \left( H_{\mu\nu}(x) + 2 \nabla_{(\mu} H_{\nu)}(x)
 + 2 \nabla_\mu\nabla_\nu H(x) + \frac{1}{5} T(x) g_{\mu\nu} \right)
 Y(y), 
\ee
where $H_{\mu\nu}$ and $H_\mu$ are transverse. The $I$ index and the
summation are suppressed here, and henceforth. The gauge freedom
$h_{MN} \rightarrow h_{MN} + 2 \nabla_{(M} \xi_{N)}$ with
$\xi_\mu(x,y) = -(H_\mu + \nabla_\mu H) Y$ and $\xi_m=0$ can be used
to fix the gauge 
\be
\label{Hgauge}
 H_\mu = H = 0.
\ee
The mixed components of the metric perturbation can be decomposed as
\be
 h_{\mu m} = (Z_1 + dZ)_\mu \hat{Y}_{1m} + (Z^+_1 + dZ^+)_\mu (dY)_m +
 (Z^-_1 + dZ^-)_\mu (d^cY)_m, 
\ee
where $Z_1$ and $Z^\pm_1$ are transverse 1-forms in $AdS_5$ and
$\hat{Y}_1$ is a doubly transverse 1-form on $M_6$. 

As described above, the internal components of the metric perturbation
can be decomposed into hermitian and anti-hermitian parts, and the
hermitian part written in terms of a $(1,1)$-form: 
\be
 h_{mn}  = -X_{(1,1)mp} J^p{}_n + A_{mn}.
\ee
We decompose $X_{(1,1)}$ as described above:
\be
 X_{(1,1)} = h(x) \hat{Y}_{(1,1)}(y) + 2N^{(1,0)}  (x) \bp
 \hat{Y}_{(1,0)}(y) +  2N^{(0,1)}(x) \p \hat{Y}_{(0,1)}(y)  + Q(x)
 dd^c Y + \frac{1}{6} J_{mn} S(x) Y(y), 
\ee
where $\hat{Y}_{(1,1)}$ is primitive and transverse and
$\hat{Y}_{(1,0)}$, $\hat{Y}_{(0,1)}$ are transverse. Note that $N^{(0,1)}$ and $N^{(1,0)}$ are (complex conjugate) scalar fields in AdS. It is convenient
to suppress the indices on $N$ and write this as 
\be
 X_{(1,1)} = h \hat{Y}_{(1,1)}+ N d\hat{Y}_1 + M d^c \hat{Y}_1  +
 Qdd^c Y + \frac{1}{6} J_{mn} S Y, 
\ee
where $N d\hat{Y}_1 \equiv N^{(1,0)} d\hat{Y}_{(1,0)} + N^{(0,1)} d\hat{Y}_{(0,1)}$,  
$M d^c \hat{Y}_1 \equiv M^{(1,0)} d^c \hat{Y}_{(1,0)} + M^{(0,1)}d^c \hat{Y}_{(0,1)}$ and
\be
  M^{(1,0)}=-iN^{(1,0)}, \qquad M^{(0,1)} = iN^{(0,1)}.
\ee
We will sometimes write this as $M = \mp i N$ where the upper and lower signs refers to
$(1,0)$ or $(0,1)$ respectively.   
 
The anti-hermitian part $A_{mn}$  can be decomposed as in
(\ref{eqn:antihermitian}): 
\be
 A_{mn}(x,y) = A(x) \hat{Y}_{Tmn}(y) + B^{(1,0)}(x) \nabla^+_{(m}
 Y_{n)(1,0)}(y) + B^{(0,1)}(x)\nabla^-_{(m} Y_{n)(0,1)}(y), 
\ee
where $\hat{Y}_{Tmn}$ denotes a transverse anti-hermitian tensor
eigenfunction of the Lichnerowicz operator on $M_6$: 
\be
 \Delta_L Y_{Tmn} \equiv -\nabla^2 Y_{Tmn} - 2 R_{mpnq}Y_T^{pq} + 4c^2
 Y_{Tmn} =  \lambda_T Y_T. 
\ee
A gauge transformation with $\xi_\mu=0$ and $\xi_m = -(1/2)( B^{(1,0)}
Y_{m(1,0)} + B^{(0,1)} Y_{m(0,1)})$ can be used to set 
\be
 B^{(1,0)} = B^{(0,1)} = 0.
\ee
Note that this gauge transformation preserves (\ref{Hgauge}).
There is some residual gauge freedom:
\be
\label{eqn:residualgauge}
\xi_\mu = k_\mu(x) Y(y), \qquad \xi_m = \alpha(x) V_m(y),
\ee
where $k_\mu$ and $V_m$ are Killing vector fields in $AdS_5$ and $M_6$
respectively. As discussed above, the latter can always be written in
terms of scalar harmonics \cite{besse,pope} 
\be
 V_m = (d^c Y)_m, \qquad \Delta_6 Y = 4c^2 Y.
\ee
The decomposition of the 3-form is:
\bea
 a &=& j Y_{(3)} + k^+ d Y_{(2)} + k^- d^c Y_{(2)} + (p_1 + dp)
 \wedge Y_{(2)} + \ell d d^c Y_{(1)} + m J \wedge Y_{(1)} \nn 
 &+& (q^+_1 + dq^+ ) \wedge d Y_{(1)} +  (q^-_1 + dq^- )
 \wedge d^c Y_{(1)} + (t_2 + dt_1) \wedge Y_{(1)} \nn 
 &+& (r_1 + dr ) \wedge d d^c Y + (u_2^+ + du_1^+ ) \wedge
 d Y + (u_2^- + du_1^- ) \wedge d^c Y \nn 
 &+& n^+ J \wedge dY + n^- J \wedge d^c Y + (s_1 + ds ) \wedge J Y
 + (w_3 +  dw_2 ) Y. 
\eea
We remind the reader that a sum over harmonics is understood, i.e., $j
Y_{(3)}$ stands for $j^I(x) Y^I_{(3)}(y)$. $j$, $k^\pm$ etc are
scalars in $AdS_5$, $p_1$, $q_1^\pm$ etc are transverse vectors
in $AdS_5$, $u_2^\pm$ etc are transverse 2-forms in $AdS_5$. We are also using the shorthand notation introduced above, e.g., $t_2 \wedge Y_1$ stands for $t_2^{(1,0)} \wedge Y_{(1,0)} + t_2^{(0,1)} \wedge Y_{(0,1)}$ where $t_2^{(1,0)}$ and $t_2^{(0,1)}$ are complex conjugate 2-forms.
In the final term, it will sometimes be convenient to rewrite
the transverse 3-form $w_3$ in terms of a transverse 1-form $v_1$: 
\be
 w_3= \star_5 d v_1.
\ee 
The 3-form $a$ has gauge freedom $a \rightarrow a+d\Lambda$ for some 2-form $\Lambda$. However, the quantities in the
above decomposition must arrange themselves into gauge-invariant
combinations when we calculate the 4-form $f$.  
Computing $f$ reveals that there is no loss of generality in imposing
the gauge conditions 
\be
 u_1^+ = q^+ = r = p = \ell = s = t_1=r_1= w_2 = 0.
\ee
We then have
\bea
 f&=& jdY_3+ k^- dd^c Y_2 + m J \wedge dY_1 + n^- J \wedge d d^c Y \nn
  &+&dj \wedge Y_3 - (p_1 - dk^+) \wedge dY_2 +dk^- \wedge d^c Y_2 -
 (q_1^- + dq^-) \wedge d d^c Y_1 \nn  
  &+& dm \wedge J \wedge Y_1 - (s_1 - dn^+) \wedge J \wedge dY + dn^-
 \wedge J \wedge d^c Y \nn 
 &+&   dp_1 \wedge Y_2 + (t_2 + dq_1^+) \wedge dY_1 + dq_1^- \wedge
 d^c Y_1 + (u_2^- + du_1^- ) \wedge d d^c Y + ds_1 \wedge J Y \nn 
 &+& dt_2 \wedge Y_1 + (-w_3 + d u_2^+ ) \wedge dY + du_2^- \wedge d^c Y \nn
 &+& dw_3 Y
 \eea
There is some ambiguity in the decomposition of the $AdS_5$ fields
into a transverse part and an exact part. An expression of the form
$V_p + dV_{p-1}$, where $V_p$ and $V_{p-1}$ are transverse forms in
$AdS_5$, is invariant under  
\be
\label{eqn:residual}
 V_{p-1}  \rightarrow V_{p-1}  + \delta V_{p-1} , \qquad V_p
 \rightarrow V_p - d \delta V_{p-1}
\ee 
where $\delta V_{p-1}$ is transverse and satisfies the equation of motion of a massless field in $AdS_5$:
\be
  \Delta \delta V_{p-1}  =0.
\ee

\subsection{The Maxwell equation}

Perturbing the Maxwell equation gives
\be
 \star d \star f + \star d \delta(\star) \bar{F} = \star \left( f
 \wedge \bar{F} \right), 
\ee
where a bar refers to the unperturbed solution and
$\delta(\star) \bar{F}$ denotes the change in $\star \bar{F}$ resulting from the metric perturbation.
In evaluating this equation, the following results are useful. Let
$X_p$ and $Y_q$ denote a $p$-form in $AdS_5$ and a $q$-form in $M_6$
respectively. Then 
\be
 \star (X_p \wedge Y_q) = (-)^{pq} (\star_5 X_p) \wedge (\star_6 Y_q),
\ee
Now take $q=4-p$ with $X_p \wedge Y_{4-p}$ a typical term in the
decomposition of the Maxwell perturbation $f$. On the LHS of the
Maxwell equation we will encounter terms of the form 
\be
\star d \star (X_p \wedge Y_{4-p}) = - \left( d_5^\dagger X_p \right)
\wedge Y_{4-p} + (-)^{p+1} X_p \wedge d_6^\dagger Y_{4-p}. 
\ee
The metric perturbation also enters the LHS of the Maxwell equation. We find
\bea
 \star d \delta(\star) F& = &-c J \wedge d_6^c h^M_M + 4c d_6^c
 X_{(1,1)} + cJ \wedge d_6^\dagger X_{(1,1)}\nn  &-& 2 c d_5^\dagger X_1' J
 \wedge J \cdot Y_1' + 2c X_1' \wedge d_6^c (J \cdot Y_1') + 2c X_1'
 \wedge J \wedge d_6^\dagger (J \cdot Y_1'), 
\eea
where $X_1'$, $Y_1'$ denote the various terms arising from the mixed
components $h_{\mu m}$, i.e., $h_{\mu m}$ is a sum of terms of the
form $(X_1')_\mu (Y_1')_m$, and the corresponding sum should be
understood in the above expression. 

Using these results, the Maxwell equation decomposes as follows. The
$\mu\nu\rho$ components give 
\be
\label{eqn:ads3formpart}
 \lambda \star d u_2^+ + 2 c \lambda  u_2^- + d \left[
   (\Delta+\lambda) v_1 + 2c\lambda u_1^-   + 6 c  s_1\right] = 0. 
\ee
The $\mu\nu m$ components describe 1-forms on $M_6$. These can be
decomposed into a transverse 1-form part, arising from terms
proportional to $\hat{Y}_{1m}$ and scalar parts proportional to $dY$
and $d^c Y$ respectively. The transverse $(1,0)$-form part is ($t_2$
denotes $t_2^{(1,0)}$ etc)  
\be  
(\Delta + \lambda_1) t_2 + \lambda_1
dq_1^+ + 2ic \star d t_2 = 0.   
\ee  
The transverse $(0,1)$-form part
is the complex conjugate of this. Now $\lambda_1 \ne 0$ (see above) so
acting on this equation with $d^\dagger$ gives $\Delta q_1^+=0$. This
implies that $q_1^+$ can be gauged away using the freedom
(\ref{eqn:residual}), i.e., we can absorb $q_1^+$ into $t_2$. So we
   set $q_1^+ = 0$ henceforth. 
This leaves 
\be 
\label{eqn:t2eq}
(\Delta + \lambda_1) t_2 + 2ic \star d t_2 = 0.  
\ee 
The
terms proportional to $dY$ give the same equation as $d$ acting on
(\ref{eqn:ads3formpart}), while the terms proportional to $d^c Y$ give
\be 
(\Delta + \lambda) u_2^- - 2c \star du_2^+ + d\left[ \lambda u_1^- +
   s_1 - 2c v_1\right] =0, \qquad \lambda \ne 0.   
\ee 
The restriction $\lambda \ne 0$ arises because otherwise $d^c Y =0$. 
The 1-form and 2-form parts of this equation and equation
   (\ref{eqn:ads3formpart}) can be decoupled using the gauge freedom
   (\ref{eqn:residual}). Consider a transformation $u_1^- \rightarrow
   u_1^- + \delta u_1^-$, $v_1 \rightarrow v_1 + \delta v_1$, $u_2^-
   \rightarrow u_2^- - d \delta u_1^-$, $u_2^+ \rightarrow u_2^+ +
   \delta u_2^+$, with $\Delta \delta v_1 = \Delta \delta u_1^- = 0$
   and $\delta u_2^+$ is defined by $d \delta u_2^+ = \star d \delta
   v_1$. This leaves the 4-form invariant. Acting with $d^\dagger$ on
   the above equations implies that the square brackets in both are
   annihilated by $\Delta$. This implies that we can choose $\delta
   u_1^-$ and $\delta v_1$ to make these brackets vanish. Hence the
   2-form and 1-form parts of these equations decouple. The 2-form
   equations give 
\be
\label{eqn:u2plus} 
  \star d u_2^+ + 2 c  u_2^-=0, \qquad \lambda \ne 0,
\ee
and (after using this equation to eliminate $u_2^+$),
\be
\label{eqn:u2minus}
 (\Delta + \lambda + 4c^2) u_2^-  =0, \qquad \lambda \ne 0.
\ee
Hence $u_2^-$ is a massive 2-form field with $m^2 = \lambda + 4c^2$,
   and $u_2^+$ is not independent. If $\lambda=0$ then $u_2^\pm$ drop
   out of the expression for $f$ and are therefore unphysical. The
   1-form equations are 
\be
\label{eqn:11eq1}
  (\Delta+\lambda) v_1 + 2c\lambda u_1^-   + 6 c  s_1 = 0,
\ee
\be
\label{eqn:11const1}
 \lambda u_1^- + s_1 - 2c v_1 = 0, \qquad \lambda \ne 0.
\ee
The $\mu m n$ components of the Maxwell equation correspond to 2-forms
   on $M_6$, which can be decomposed into irreducible pieces as
   described above. The terms proportional to $\hat{Y}_2$ give $\Delta
   p_1 + \lambda_2  ( p_1 - dk^+) = 0$, which implies that $\Delta
   k^+=0$ so we can absorb $k^+$ into $p_1$ using the residual freedom
   (\ref{eqn:residual}). This leaves 
\be
\label{eqn:p1eq}
 \left (\Delta + \lambda_2 \right) p_1 = 0.
\ee
Hence $p_1$ is a vector field with $m^2 = \lambda_2$. The terms
proportional to $d\hat{Y}_1$ vanish when we use $q_1^+=0$. The terms
proportional to $d^c \hat{Y}_1$ give, for a $(1,0)$-form $\hat{Y}_1$
(so $q_1^-$ denotes $q_1^{-(1,0)}$ etc, $(0,1)$-forms give the complex
conjugate of this) a 1-form part\footnote{The split into 1-form and
scalar parts uses the freedom (\ref{eqn:residual}) as described
above. Strictly speaking, this can only be done once we have the
complete set of equations governing these fields, but we shall
anticipate the final result and split the equations as we encounter them.}
\be
\label{eqn:q1minusZ11}
 \left( \Delta   + \lambda_1 \right) q_1^- + 2ic Z_1  = 0,
\ee
and a scalar part
\be
\label{eqn:01const1}
  \lambda_1 q^- - m +2ic Z = 0.
\ee
The terms proportional to $dd^c Y$ give (NB $dd^cY=0$ if, and only if,
$\lambda=0$) a 1-form part
\be
\label{eqn:11eq2}
\Delta u_1^- -  s_1 - 2c Z_1^-  = 0 \qquad \lambda \ne 0,
\ee
and a scalar part
\be
\label{eqn:nplus}
 n^+ = 2c Z^- \qquad \lambda \ne 0.
\ee
The terms proportional to $J Y$ give 1-form part
\be
\label{eqn:11eq3}
 \Delta s_1 + \lambda s_1 + 2c \lambda Z_1^-  - 2c \Delta v_1 = 0.
\ee
and the scalar part reproduces (\ref{eqn:nplus}).

Finally, we consider the $mnp$ components of the Maxwell
equation. These transform as a 3-form on $M_6$, which can be
decomposed as described above. Doing so, the terms proportional to
$\hat{Y}_3$ give
\be
\label{eqn:jeq}
 (\Delta + \lambda_3) j = 0,
\ee
so the scalar field $j$ has $m^2 = \lambda_3$. The terms proportional
to $d\hat{Y}_2$ vanish (using $k^+=0$). Terms proportional to $d^c
\hat{Y}_2$ give (if $\lambda_2=0$ then $\hat{Y}_2$ is harmonic so $d^c \hat{Y}_2=0$)
\be
\label{eqn:kminusH1}
 (\Delta + \lambda_2) k^- - 4 c h = 0 \qquad \lambda_2 \ne 0.
\ee
Terms proportional to $dd^c \hat{Y}_1$ give
\be
\label{eqn:01eq1}
 \Delta q^- +m - 4cN = 0.
\ee
Terms proportional to $J \wedge \hat{Y}_1$ give (this comes from the $(1,0)$
part of $\hat{Y}_1$, the $(0,1)$ part gives the complex conjugate) 
\be
\label{eqn:01eq2}
 (\Delta + \lambda_1 )m - 4 c \lambda_1 N - 2 i c \Delta Z = 0,
\ee
Terms proportional to $J \wedge dY$ vanish upon using
(\ref{eqn:nplus}). Terms proportional to $J \wedge d^c Y$ give
\be
\label{eqn:00eq1}
 (\Delta + \lambda) n^- + cT - \frac{1}{3} cS - 2 c \lambda Q+ 2 c
\Delta Z^+ = 0 \qquad \lambda \ne 0.
\ee
 
\subsection{The Einstein equation}

The perturbed Einstein equation is
\be
 \delta R_{MN} = \delta S_{MN},
\ee
where
\be
 \delta R_{MN} = -\frac{1}{2} \left( \nabla_5^2 + \nabla_6^2 \right)
 h_{MN} - \frac{1}{2} \nabla_M \nabla_N h^P_P + \nabla_{(M} \nabla^P
 h_{N)P} - \bar{R}_{MPNQ} h^{PQ} + \bar{R}_{(M}^P h_{N)P},
\ee
and
\bea
 \delta S_{MN} &=& \frac{1}{12} \left[2 f_{(M|PQR|} \bar{F}_{N)}{}^{PQR} - 3
   \bar{F}_{MPRS} \bar{F}_{NQ}{}^{RS} h^{PQ} \right. \nn &-&
   \left. \frac{1}{12} h_{MN} 
   \bar{F}_{PQRS} \bar{F}^{PQRS} - \frac{1}{12} \bar{g}_{MN} \left( 2
   f_{PQRS} \bar{F}^{PQRS} - 4 \bar{F}_{PRST}\bar{F}_Q{}^{RST} h^{PQ}
   \right) \right].
\eea
Evaluating the $\mu\nu$ components and decomposing into irreducible parts
gives transverse traceless tensor part
\be
\label{eqn:spin2}
 - \nabla_5^2 H_{\mu\nu} + (\lambda-c^2) H_{\mu\nu}=0,
\ee
so $H_{\mu\nu}$ is a massive spin-2 field, for $\lambda=0$ we obtain
the massless $AdS_5$ graviton. The 1-form part is
\be
\label{eqn:Z1plus}
 \nabla_{(\mu} Z^+_{1\nu)} = 0 \qquad \lambda \ne 0,
\ee
which implies that $Z^+_1$ can be gauged away using the residual gauge
invariance (\ref{eqn:residualgauge}). (If $\lambda=0$ then $Z^+_1$
drops out of $h_{\mu m}$ so is unphysical.) Hence we set $Z^+_1 = 0$
henceforth. Terms of the form $\nabla_\mu \nabla_\nu (\rm scalar)$
give
\be
\label{eqn:00const1}
 \lambda Q + \frac{1}{2} S + \frac{3}{10} T + \lambda Z^+ = 0,
\ee
and terms proportional to $\bar{g}_{\mu\nu}$ give
\be
\label{eqn:00eq2}
 \frac{1}{10} \left( \Delta + \lambda + 4c^2 \right ) T + \frac{4}{3}
 \left( c \lambda n^- - c^2 S - 2 c^2 \lambda Q \right) = 0.
\ee
The $\mu m$ components of the Einstein equation can be decomposed into
transverse 1-form and scalar parts on $M_6$. These can then be
decomposed into transverse 1-form and scalar parts on $AdS_5$.
The transverse $(1,0)$-form part gives $AdS_5$ 1-form equation
\be
\label{eqn:q1minusZ12}
 \frac{1}{2} (\Delta + \lambda_1+4c^2) Z_1 - ic \lambda_1 q_1^- =0,
\ee
and the $AdS_5$ scalar part is
\be
\label{eqn:01const2}
 \frac{1}{2} \lambda_1 (Z+iN - 2ic  q^-) + 2ic m +2c^2 Z = 0.
\ee
From terms proportional to $dY$ we obtain vanishing $AdS_5$ 1-form
part (using $Z_1^+=0$). The scalar part is
\be
\label{eqn:00const2}
 \lambda Q + \frac{5}{6} S + \frac{4}{5}T - 4c n^- - 4c^2 Z^+ = 0
 \qquad \lambda \ne 0.
\ee
From terms proportional to $d^c Y$ we obtain 1-form part
\be
\label{eqn:11eq4}
 \left( \Delta + \lambda + 4c^2 \right) Z_1^- + 4 c s_1 = 0 \qquad
 \lambda \ne 0,
\ee
and scalar part
\be
 (\lambda + 4c^2) Z^- = 4c n^+ \qquad \lambda \ne 0.
\ee
Combining this with (\ref{eqn:nplus}) gives
\be
\label{eqn:nplusZminus}
 n^+ = Z^- = 0,
\ee
unless $\lambda=0$ or $\lambda=4c^2$. In the former case, $n^+$ and
$Z^-$ are unphysical. The latter case corresponds to a harmonic $Y$
for which $d^c Y$ is a Killing field on $M_6$. In this case, we can
use the residual gauge freedom (\ref{eqn:residualgauge}) to set $Z^- =
0$ so equation (\ref{eqn:nplus}) gives $n^+ = 0$. Hence equation
(\ref{eqn:nplusZminus}) is satisfied in general.

Next consider the $mn$ components of the Einstein equation, which only
involve $AdS_5$ scalars. First we decompose these into hermitian and
anti-hermitian parts. The transverse anti-hermitian part gives
\be
\label{eqn:Aeq}
 \left( \Delta + \lambda_T - 4c^2 \right) A = 0,
\ee
where $\lambda_T$ is an eigenvalue of the Lichnerowicz operator on
$M_6$ corresponding to tranverse anti-hermitian modes. The
anti-hermitian part also has transverse 1-form, and scalar parts. The
transverse $(1,0)$-form part is
\be
\label{eqn:01eq3}
 \Delta Z - i \lambda_1 N = 0.
\ee
After using $Z^-=0$, the scalar part, proportional to $\nabla^\pm _m
\nabla^\pm _n Y$ gives
\be
\label{eqn:00eq3}
 \Delta Z^+ + \frac{1}{3}S + \frac{1}{2} T  = 0 \qquad \lambda \ne 0.
\ee
The hermitian part of the $mn$ Einstein equation can be converted to a
$(1,1)$-form and decomposed as described above. The transverse
primitive part gives
\be
\label{eqn:kminusH2}
 \left( \Delta + \lambda_{(1,1)} + 4c^2 \right) h^{(1,1)} - 4 c
 \lambda_{(1,1)} k^{-(1,1)} = 0.
\ee
The transverse vector part gives
\be
\label{eqn:01eq4}
 \left( \Delta + \lambda_1 + 4c^2 \right) N -2 c m = 0.
\ee
The scalar part proportional to $dd^c Y$ gives
\be
\label{eqn:00eq4}
 \left( \Delta + \lambda + 4c^2 \right) Q - 4cn^- = 0 \qquad \lambda
 \ne 0.
\ee
The scalar part proportional to $JY$ gives
\be
\label{eqn:00eq5}
 \left( \Delta + \lambda + 12c^2 \right) S - 8 c \lambda n^- + 16 c^2
 \lambda Q = 0.
\ee 

\subsection{The mass spectrum}

In this section we shall diagonalize the above equations to determine
the full Kaluza-Klein spectrum.

\subsubsection{Symmetric tensor/scalar modes}\label{sec:spin-2}

This sector contains just the real, transverse, traceless, symmetric
tensor field $H_{\mu\nu}$ with equation of motion
(\ref{eqn:spin2}). For $\lambda=0$ this gives the $AdS_5$
graviton. $\lambda>0$ gives massive spin-2 fields.

\subsubsection{2-form/1-form modes}\label{sec:2f-1f}

In this sector we have the complex field $t_2$. The equation of
motion is (\ref{eqn:t2eq}). To
obtain the mass associated with this field, we note that a complex
2-form in $AdS_5$ has a {\it first} order equation of motion
\cite{arutyunov}, so $t_2$ is actually equivalent to two complex 2-form
fields. Equation (\ref{eqn:t2eq}) can be decomposed into first order
equations by defining
\be
Z_2 = t_2 + i a \star_5 dt_2,
\ee
and seek $a$ so that $\star_5 d Z_{2} \propto Z_{2}$.
This requires $\lambda_1 a^2 + 2ac - 1=0$, so there are two solutions:
$\lambda_1 a_\pm = \mp \sqrt{\lambda_1 + c^2} - c$. Hence
there are two linearly independent solutions $Z_{2}^\pm$. Obviously
$t_2$ can be written as a linear combination of these two fields.
We then have 
\be
 \star d Z_{(2)}^\pm = - i a_\pm \lambda_{1} Z_{2}^\pm.
\ee
This is the equation of motion of a complex 2-form with mass given by
$m_{\pm}^2 = (a_\pm \lambda_1)^2$ (see e.g. \cite{arutyunov}). To see this, note
that acting with $\star d$ gives
\be
(\Delta_5 +
(a_\pm\lambda_1)^2) Z_{2}^\pm = 0.
\ee
Hence we have two complex 2-form fields of definite mass, namely
$Z_{(2)}^\pm$, with masses given by  
\be
 m_\pm = \sqrt{\lambda_1 + c^2} \pm c .
\ee
As discussed above, $\lambda_1>0$ so these fields are both massive.

\subsubsection{2-form/scalar modes}\label{sec:2f-s}

In this section we have the real fields $u_2^\pm$. $u_2^+$ is given in
terms of $u_2^-$ by equation (\ref{eqn:u2plus}), and $u_2^-$ has equation of motion
(\ref{eqn:u2minus}). Hence this sector contains a single real 2-form
with $m^2 = \lambda+4c^2$, $\lambda>0$.

\subsubsection{1-form/2-form modes}\label{sec:1f-2f}

The only field in this sector is $p_1$, with equation of motion
(\ref{eqn:p1eq}). This can be decomposed into
the complex field $p_1^{(2,0)}$ (with complex conjugate $p_1^{(0,2)}$)
with $m^2 = \lambda_{(2,0)}$ 
and a real field $p_1^{(1,1)}$ (since we can take $(1,1)$-form
eigenfunctions of $\Delta_6$ to be real) with $m^2 =
\lambda_{(1,1)}$. Note that (primitive, transverse) harmonic 2-forms give rise to
massless 1-forms in $AdS_5$.

\subsubsection{1-form/1-form modes}\label{sec:1f-1f}

In this sector we have the complex 1-form fields $q_1^-$ and $Z_1$ (or, more
precisely, $q_1^{-(1,0)}$ and $Z_1^{(1,0)}$) with equations of motion
(\ref{eqn:q1minusZ11}, \ref{eqn:q1minusZ12}). (We saw above that
$q_1^+$ can be gauged away.) Diagonalizing gives the
masses as
\be
 m^2 = \lambda_1 + 2c^2 \pm \sqrt{(\lambda_1 + 2c^2)^2 -\lambda_1^2}.
\ee
These fields are all massive (because $\lambda_1>0$).

\subsubsection{1-form/scalar modes}\label{sec:1f-s}

The fields in this sector are $v_1$, $u_1^-$, $s_1$ and
$Z_1^-$. (We saw above that equation (\ref{eqn:Z1plus}) implies that
$Z_1^+$ can be gauged away.) These fields are real. They are governed
by the equations of motion
(\ref{eqn:11eq1},\ref{eqn:11eq2},\ref{eqn:11eq3},\ref{eqn:11eq4}) and
the constraint (\ref{eqn:11const1}).

Consider first the case $\lambda=0$. In this case, the only physical
fields are $s_1$ and $v_1$ and the only non-trivial equations are
 (\ref{eqn:11eq1}), which gives $\Delta v_1 + 6 c s_1 = 0$, and
(\ref{eqn:11eq3}), which gives $\Delta (s_1 -2 cv_1) = 0$. Combining
these gives
\be
 \Delta (s_1 - 2 c v_1 ) = 0, \qquad \left( \Delta + 12 c^2 \right)
 s_1 = 0, \qquad \lambda=0.
\ee
Hence $s_1 -2 cv_1$ is massless and $s_1$ has $m^2 = 12c^2$. Recall
that $v_1$ arises from the $AdS_5$ components of the 
M-theory 3-form via $w_3 = \star d v_1$. Hence the massless field we
have found here is essentially the Kaluza-Klein zero mode of the
M-theory 3-form. This massless 3-form can be dualized to a scalar via
$d (w_3 - (1/2c) \star d s_1)  = \star d \sigma$. This scalar has a
gauge invariance $\sigma \sim \sigma + {\rm constant}$.

Now consider the case $\lambda \ne 0$. It can be verified that the
constraint equation (\ref{eqn:11const1}) is consistent with the four
equations of motion. This constraint can be used to eliminate, say,
$s_1$. This leaves three fields. The equations of motion can be combined to give
\be
 \left( \Delta + \lambda + 4c^2 \right) (v_1 - Z_1^-) = 0 \qquad
 \lambda \ne 0,
\ee
so $v_1 -Z_1^-$ is a field with $m^2 = \lambda + 4c^2$. The remaining
two mass eigenstates can be identified by setting ${\cal U}_1 = u_1^- + \alpha v_1 +
\beta (Z_1^- - v_1)$ and choosing $\alpha$, $\beta$ so that $(\Delta
+ m^2) {\cal U}_1=0$ for some $m$. This gives
$\beta = 1/(2\lambda \alpha + 2c)$, $ \alpha = (-3c
   \mp \sqrt{9 c^2 + 4 \lambda})/(2\lambda)$. Denote the corresponding
   linear combinations as ${\cal U}_{1\pm}$. Their masses
are
\be
 m_\pm^2 = \lambda + 6c^2 \pm \sqrt{(\lambda+6c^2)^2 -
   \lambda(\lambda-4c^2)} \qquad \lambda \ne 0.
\ee
Hence, for $\lambda=4c^2$, ${\cal U}_{1-}$ is a massless real vector
field. But scalar modes with $\lambda=4c^2$ are in one-to-one
correspondence with Killing vector fields on $M_6$. Hence these
massless vectors must be the Kaluza-Klein gauge bosons.

\subsubsection{Scalar/anti-hermitian tensor modes}\label{sec:s-ah}

A symmetric anti-hermitian tensor can be decomposed into $(2,0)$ and
$(0,2)$ parts, so we have two complex conjugate fields $A^{(2,0)}$ and
$A^{(0,2)}$, with equation of motion given by (\ref{eqn:Aeq}). Hence we
have $m^2 = \lambda_T -4c^2$. This can be seen to be non-negative
using the following standard argument that relates anti-hermitian eigenfunctions
of the Lichnerowicz operator to complex structure deformations \cite{besse}. 

Consider an anti-hermitian $(2,0)$ tensor eigenfunction $\hat{Y}_{mn}$ with
eigenvalue $\lambda$. Raising an index, we have a tensor $\hat{Y}^m_n$
which can be regarded as a 
$(0,1)$-form taking values in $T^{1,0} M_6$, the holomorphic tangent
space of $M_6$. For a $(0,q)$-form $\omega$ taking values in $T^{1,0}
M_6$ we define 
\be
 (\bp \omega)^m_{n p_1 \ldots p_q} = (q+1) \nabla^-_{[n} \omega^m_{p_1 \ldots p_q]},
\ee
where $\nabla^-_m$ denote the $(0,1)$ part of $\nabla_m$. For any two
such forms $\omega$, $\nu$ we define the obvious inner product 
\be
 (\omega,\nu) = \frac{1}{q!} \int \omega^m_{n_1 \ldots n_p} \, g_{mm'}
g^{n_1 n_1'} \ldots g^{n_p n_p'} \, \bar{\nu}^{m'}_{n_1'\ldots n_p'}. 
\ee
We can then defines the adjoint $\bp^\dagger$. Transversality implies
that $(\bp^\dagger Y)^m = 0$. Now define the Laplacian acting on
$(0,q)$-forms taking values in $T^{1,0} M_6$ by $\Delta_{\bp} \equiv
2\left( \bp \bp^\dagger + \bp^\dagger \bp \right)$. Acting on $Y$, we find that 
\be
  \left( \Delta_{\bp} Y \right)^m_n = [ (\Delta_L -4c^2) Y ]^m_n =
 (\lambda_T-4c^2) Y^m_n. 
\ee
Hence the mass of the complex scalar in this sector is given by
\be
 m^2 =  \lambda_{(0,1)}^{(1,0)},
\ee
where $\lambda_{(0,1)}^{(1,0)}$ denotes the eigenvalues of
$\Delta_{\bp}$. These are manifestly non-negative. Modes with $m=0$
correspond to infinitesimal deformations of the complex structure of $M_6$.

\subsubsection{Scalar/3-form modes}\label{sec:s-3f}

The only field here is $j$, or, more precisely, the complex scalar $j^{(2,1)}$. 
The equation of motion is (\ref{eqn:jeq}) so $j^{(2,1)}$ has $m^2 = \lambda_{(2,1)}$.
There are no transverse $(3,0)$-forms hence there
is no $j^{(3,0)}$ part.

\subsubsection{Scalar/2-form modes}\label{sec:s-2f}

The fields in this sector are $h$ and $k^-$. Their equations of motion
are given by equations (\ref{eqn:kminusH1}) and (\ref{eqn:kminusH2}). 
Now $h$ is associated with $(1,1)$-forms, i.e., $h^{(2,0)} =
h^{(0,2)}=0$. Hence (\ref{eqn:kminusH1}) gives 
\be
  (\Delta + \lambda_{(2,0)}) k^{-(2,0)} = 0 \qquad \lambda_{(2,0)} \ne
0,
\ee
and $k^{-(0,2)}$ is the complex conjugate of $k^{-(2,0)}$. So
$k^{-(2,0)}$ is a complex massive scalar field with $m^2 = \lambda_{(2,0)}>0$. 

For the $(1,1)$-forms, we have to diagonalize equations
(\ref{eqn:kminusH1}) and (\ref{eqn:kminusH2}), which was discussed
in section \ref{sec:s-1,1f}.

\subsubsection{Scalar/1-form modes}\label{sec:s-1f}

In this sector we have the complex fields $m$, $q^-$, $Z$ and
$N$. (More precisely: $m^{(1,0)}$, $q^{-(1,0)}$ etc.) These satisfy
the equations of motion
(\ref{eqn:01eq1},\ref{eqn:01eq2},\ref{eqn:01eq3},\ref{eqn:01eq4}) and
the constraints (\ref{eqn:01const1},\ref{eqn:01const2}). These
constraints are compatible with the equations of motion and can be
used to eliminate, say, $q^-$ and $Z$, leaving two fields $m$, $N$.
The equations of motion for $m$ and $N$ are (\ref{eqn:01eq4}) and
\be
 (\Delta + \lambda_1 )m - 2 c \lambda_1 N = 0.
\ee
Diagonalizing gives the masses as
\be
 m^2 = \lambda_1 + 2c^2 \pm \sqrt{(\lambda_1 + 2c^2) - \lambda_1^2}.
\ee
Since $\lambda_1>0$, these two fields are massive.

\subsubsection{Scalar/scalar modes}\label{sec:s-s}

This sector contains the real fields $n^-$, $S$, $Z^+$, $Q$, $T$ (we
saw above that $n^+ = Z^- = 0$). The equations of motion are
(\ref{eqn:00eq1},\ref{eqn:00eq2},\ref{eqn:00eq3},\ref{eqn:00eq4},\ref{eqn:00eq5})
and there are two constraints
(\ref{eqn:00const1},\ref{eqn:00const2}). It can be checked that the
constraints are consistent with the equations of motion.

If $\lambda=0$ then the only
physical modes are $S$ and $T$, obeying the equations of motion
(\ref{eqn:00eq2}, \ref{eqn:00eq5}) and the constraint
(\ref{eqn:00const1}). The constraint can be used to eliminate, $T$,
leaving
\be
 (\Delta + 12c^2 ) S =0 \qquad \lambda =0,
\ee
so for $\lambda=0$ we have a single field with $m^2 = 12c^2$.

Now assume $\lambda > 0$. The constraints can be used to eliminate $S$
and $T$, leaving three fields. The other equations can be rearranged to give
\be
 (\Delta + \lambda + 4c^2 ) (Q+Z^+) = 0 \qquad \lambda \ne 0
\ee
hence $Q+Z^+$ is a field with $m^2 = \lambda + 4c^2$. The remaining
two linear combinations with definite mass can be identified by
setting ${\cal V} = n^- + \alpha Z^+ + \beta (Q+Z^+)$ and choosing
$\alpha$, $\beta$ so that the equations of motion imply $(\Delta +
m^2) {\cal V}=0$. This requires $\beta = \lambda/(3\alpha + 3c)$ and
$\alpha = (1/2)(-c \pm \sqrt{4\lambda + 9c^2})$, corresponding to two
linear combinations ${\cal V}_\pm$. The masses are given by
\be
 m_{\pm}^2 =  \lambda+6c^2 \pm \sqrt{(\lambda+6c^2)^2 -
   \lambda(\lambda-4c^2)} \qquad \lambda \ne 0.
\ee
Scalar modes with $\lambda=4c^2$ give a massless field ${\cal
  V}_-$. As discussed above, such modes are in one-to-one
correspondence with Killing vector fields of $M_6$.

\bigskip

\begin{center}{\bf Acknowledgments}\end{center}

\medskip

\noindent We are grateful to James Sparks for very helpful discussions on properties of K\"ahler-Einstein spaces. We have also had useful discussions with Micha Berkooz, Joe Conlon, Gary Gibbons, Anshuman Maharana, Chris Pope and David Tong. JEM is supported by the University of Nottingham. HSR is a Royal Society University Research Fellow.

\appendix 

\section{Conventions}

We use a positive signature metric. The bosonic action for
eleven-dimensional supergravity is given by 
\be
 16 \pi G S = \int d^{11} x \sqrt{-g} R + \int \left( - \frac{1}{2} F
 \wedge \star F + \frac{1}{6} A \wedge F \wedge F \right), 
\ee
where $F=dA$ is the 4-form. The equations of motion are
\be
 R_{MN} = \frac{1}{12} \left( F_{MPQR} F_N{}^{PQR} - \frac{1}{12}
 g_{MN} F_{PQRS} F^{PQRS} \right), \qquad d \star F = \frac{1}{2} F
 \wedge F. 
\ee
The orientation is fixed by specifying the 11d volume form
\be
 \eta_{11} = \eta_5 \wedge \eta_6,
\ee
where $\eta_5$ and $\eta_6$ are the volume forms of $AdS_5$ and $M_6$
respectively. $\eta_6$ is related to the K\"ahler form by 
\be
 \eta_6 = 6 J \wedge J \wedge J.
\ee
On $M_6$ we have
\be
 d_6^\dagger = \star_6 d_6 \star_6,
\ee
and the Laplacian is
\be
 \Delta_6 = d_6 d_6^\dagger + d_6^\dagger d_6.
\ee
We also have the Dolbeault operators $\partial$, $\bar{\partial}$ such
that $d_6 = \partial+\bar{\partial}$. We can define an exterior
derivative $d_6^c$ using $J_m{}^n \nabla_n$, or, equivalently,
\be
 d_6^c = -i (\partial - \bar{\partial}).
\ee
On $AdS_5$, for a $p$-form $X$, we define
\be
 d_5^\dagger X_p = (-)^{p+1} \star_5 d_5 \star_5 X_p
\ee
and the wave operator is
\be
 \Delta =  d_5 d_5^\dagger + d_5^\dagger d_5.
\ee
A free $p$-form field of mass $m$ has equation of motion
\be
 \left( \Delta + m^2 \right) X_p = 0.
\ee


\begin{thebibliography}{99}

\bibitem{kkreview}
M.~J.~Duff, B.~E.~W.~Nilsson and C.~N.~Pope,
  ``Kaluza-Klein Supergravity,''
  Phys.\ Rept.\  {\bf 130}, 1 (1986).
  %%CITATION = PRPLC,130,1;%%

\bibitem{dolan84a}
B.~Dolan,
  ``A New Solution Of D = 11 Supergravity With Internal Isometry Group SU(3) X
  SU(2) X U(1),''
  Phys.\ Lett.\  B {\bf 140}, 304 (1984).
  %%CITATION = PHLTA,B140,304;%%

\bibitem{dolan84}
B.~Dolan,
  ``The Kahler Two Form In D = 11 Supergravity,''
  Class.\ Quant.\ Grav.\  {\bf 2}, 309 (1985).
  %%CITATION = CQGRD,2,309;%%

\bibitem{tian}
G.~Tian, ``On K\"ahler-Einstein metrics on certain K\"ahler manifolds with $C(1)>0$'', Invent. Math. {\bf 89}, 
225 (1987).

\bibitem{tianyau}
G.~Tian and S.~T.~Yau,
  ``Kahler-Einstein Metrics On Complex Surfaces With $C(1) > 0$,''
  Commun.\ Math.\ Phys.\  {\bf 112} (1987) 175.
  %%CITATION = CMPHA,112,175;%%

\bibitem{popevann}
C.~N.~Pope and P.~van Nieuwenhuizen,
``Compactifications of d=11 supergravity on Kahler manifolds,''
Commun.\ Math.\ Phys.\  {\bf 122}, 281 (1989).
%%CITATION = CMPHA,122,281;%%

\bibitem{gauntlett}
J.~P.~Gauntlett, D.~Martelli, J.~Sparks and D.~Waldram,
 ``Supersymmetric AdS(5) solutions of M-theory,''
 Class.\ Quant.\ Grav.\  {\bf 21}, 4335 (2004)
 [arXiv:hep-th/0402153].
 %%CITATION = CQGRD,21,4335;%%

\bibitem{adscftreview}
  O.~Aharony, S.~S.~Gubser, J.~M.~Maldacena, H.~Ooguri and Y.~Oz,
  ``Large N field theories, string theory and gravity,''
  Phys.\ Rept.\  {\bf 323} (2000) 183
  [arXiv:hep-th/9905111].
  %%CITATION = PRPLC,323,183;%%

\bibitem{dowker}
F.~Dowker, J.~P.~Gauntlett, G.~W.~Gibbons and G.~T.~Horowitz,
  ``Nucleation of $P$-Branes and Fundamental Strings,''
  Phys.\ Rev.\  D {\bf 53}, 7115 (1996)
  [arXiv:hep-th/9512154].
  %%CITATION = PHRVA,D53,7115;%%

\bibitem{seiberg}
  N.~Seiberg and E.~Witten,
  ``The D1/D5 system and singular CFT,''
  JHEP {\bf 9904} (1999) 017
  [arXiv:hep-th/9903224].
  %%CITATION = JHEPA,9904,017;%%

\bibitem{bf}
P.~Breitenlohner and D.~Z.~Freedman,
  ``Stability In Gauged Extended Supergravity,''
  Annals Phys.\  {\bf 144}, 249 (1982),
  %%CITATION = APNYA,144,249;%%
P.~Breitenlohner and D.~Z.~Freedman,
  ``Positive Energy In Anti-De Sitter Backgrounds And Gauged Extended
  Supergravity,''
  Phys.\ Lett.\  B {\bf 115}, 197 (1982),
  %%CITATION = PHLTA,B115,197;%%
G.~W.~Gibbons, C.~M.~Hull and N.~P.~Warner,
  ``The Stability Of Gauged Supergravity,''
  Nucl.\ Phys.\  B {\bf 218}, 173 (1983),
  %%CITATION = NUPHA,B218,173;%%
L.~Mezincescu and P.~K.~Townsend,
  ``Stability At A Local Maximum In Higher Dimensional Anti-De Sitter Space And
  Applications To Supergravity,''
  Annals Phys.\  {\bf 160}, 406 (1985).
  %%CITATION = APNYA,160,406;%%

\bibitem{duffpopestability}
M.~J.~Duff, B.~E.~W.~Nilsson and C.~N.~Pope,
  ``The Criterion For Vacuum Stability In Kaluza-Klein Supergravity,''
  Phys.\ Lett.\  B {\bf 139}, 154 (1984).
  %%CITATION = PHLTA,B139,154;%%

\bibitem{freedmangubser}
O.~DeWolfe, D.~Z.~Freedman, S.~S.~Gubser, G.~T.~Horowitz and I.~Mitra,
  ``Stability of AdS(p) x M(q) compactifications without supersymmetry,''
  Phys.\ Rev.\  D {\bf 65}, 064033 (2002)
  [arXiv:hep-th/0105047].
  %%CITATION = PHRVA,D65,064033;%%

\bibitem{berkooz}
M.~Berkooz and S.~J.~Rey,
  ``Non-supersymmetric stable vacua of M-theory,''
  JHEP {\bf 9901}, 014 (1999)
  [Phys.\ Lett.\  B {\bf 449}, 68 (1999)]
  [arXiv:hep-th/9807200].
  %%CITATION = PHLTA,B449,68;%%

\bibitem{james}
J. Sparks, private comunication. 

\bibitem{horowitz}
  G.~T.~Horowitz, J.~Orgera and J.~Polchinski,
  ``Nonperturbative Instability of $AdS_5 \times S^5/Z_k$,''
  Phys.\ Rev.\  D {\bf 77}, 024004 (2008)
  [arXiv:0709.4262 [hep-th]].
  %%CITATION = PHRVA,D77,024004;%%

\bibitem{besse}
A.L. Besse, ``Einstein manifolds'', Springer, U.S.A. (2008)

\bibitem{pope}
P.~Hoxha, R.~R.~Martinez-Acosta and C.~N.~Pope,
  ``Kaluza-Klein consistency, Killing vectors, and Kaehler spaces,''
  Class.\ Quant.\ Grav.\  {\bf 17}, 4207 (2000)
  [arXiv:hep-th/0005172].
  %%CITATION = CQGRD,17,4207;%%

\bibitem{cp3eigenvalues}
A. Ikeda and Y. Taniguchi, ``Spectra and eigenforms of the laplacian
on $S^n$ and $P^n(C)$'', Osaka J. Math {\bf 15}, 515 (1978).

\bibitem{klr}
H.~K.~Kunduri, J.~Lucietti and H.~S.~Reall,
  ``Gravitational perturbations of higher dimensional rotating black holes:
  Tensor Perturbations,''
  Phys.\ Rev.\  D {\bf 74}, 084021 (2006)
  [arXiv:hep-th/0606076].
  %%CITATION = PHRVA,D74,084021;%%

\bibitem{lunin}
O.~Lunin and J.~M.~Maldacena,
  ``Deforming field theories with U(1) x U(1) global symmetry and their
  gravity duals,''
  JHEP {\bf 0505}, 033 (2005)
  [arXiv:hep-th/0502086].
  %%CITATION = JHEPA,0505,033;%%

\bibitem{arutyunov}
G.~E.~Arutyunov and S.~A.~Frolov,
 ``Antisymmetric tensor field on AdS(5),''
 Phys.\ Lett.\  B {\bf 441}, 173 (1998)
 [arXiv:hep-th/9807046].
  %%CITATION = PHLTA,B441,173;%%

\end{thebibliography}
\end{document}